\documentclass[prl,aps,twocolumn, superscriptaddress]{revtex4}
\usepackage{hyperref}
\usepackage{xcolor}
\usepackage{graphicx}
\usepackage{dcolumn}
\usepackage{bm}
\usepackage{bbm}
\usepackage{hhline} 
\usepackage{physics}

\begin{document}

\title{Qutrit randomized benchmarking}
\author{A. Morvan}
\email{amorvan@lbl.gov}
\affiliation{Quantum Nanoelectronics Laboratory, Dept. of Physics,
             University of California at Berkeley, Berkeley, CA 94720, USA}
\affiliation{Computational Research Division, Lawrence Berkeley National Lab, Berkeley, CA 94720, USA}
\author{V. V. Ramasesh}
\affiliation{Quantum Nanoelectronics Laboratory, Dept. of Physics,
             University of California at Berkeley, Berkeley, CA 94720, USA}
\author{M. S. Blok}
\affiliation{Quantum Nanoelectronics Laboratory, Dept. of Physics,
             University of California at Berkeley, Berkeley, CA 94720, USA}
\affiliation{Department of Physics and Astronomy, University of Rochester, Rochester, NY, 14627 USA}
\author{J.M. Kreikebaum}
\affiliation{Quantum Nanoelectronics Laboratory, Dept. of Physics,
             University of California at Berkeley, Berkeley, CA 94720, USA}
\affiliation{Materials Sciences Division, Lawrence Berkeley National Lab, Berkeley, CA 94720, USA}
\author{K. O'Brien}
\affiliation{Department of Electrical Engineering and Computer Science, Massachusetts Institute of Technology, Cambridge, MA 02139, United States of America}
\author{L. Chen}
\affiliation{Quantum Nanoelectronics Laboratory, Dept. of Physics,
             University of California at Berkeley, Berkeley, CA 94720, USA}
\author{B. K. Mitchell}
\affiliation{Quantum Nanoelectronics Laboratory, Dept. of Physics,
             University of California at Berkeley, Berkeley, CA 94720, USA}
\author{R. K. Naik}
\affiliation{Quantum Nanoelectronics Laboratory, Dept. of Physics,
             University of California at Berkeley, Berkeley, CA 94720, USA}
\author{D. I. Santiago}
\affiliation{Quantum Nanoelectronics Laboratory, Dept. of Physics,
             University of California at Berkeley, Berkeley, CA 94720, USA}
\affiliation{Computational Research Division, Lawrence Berkeley National Lab, Berkeley, CA 94720, USA}
\author{I. Siddiqi}
\affiliation{Quantum Nanoelectronics Laboratory, Dept. of Physics,
             University of California at Berkeley, Berkeley, CA 94720, USA}
\affiliation{Computational Research Division, Lawrence Berkeley National Lab, Berkeley, CA 94720, USA}
\affiliation{Materials Sciences Division, Lawrence Berkeley National Lab, Berkeley, CA 94720, USA}
\date{\today}

\begin{abstract}
Ternary quantum processors offer significant computational advantages over conventional qubit technologies, leveraging the encoding and processing of quantum information in qutrits (three-level systems). To evaluate and compare the performance of such emerging quantum hardware it is essential to have robust benchmarking methods suitable for a higher-dimensional Hilbert space. We demonstrate extensions of industry standard Randomized Benchmarking (RB) protocols, developed and used extensively for qubits, suitable for ternary quantum logic. Using a superconducting five-qutrit processor, we find a single-qutrit gate infidelity as low as $2.38 \times 10^{-3}$. Through interleaved RB, we find that this qutrit gate error is largely limited by the native (qubit-like) gate fidelity, and employ simultaneous RB to fully characterize cross-talk errors. Finally, we apply cycle benchmarking to a two-qutrit CSUM gate and obtain a two-qutrit process fidelity of $0.82$. Our results demonstrate a RB-based tool to characterize the obtain overall performance of a qutrit processor, and a general approach to diagnose control errors in future qudit hardware.
\end{abstract}

\maketitle

\emph{Introduction}---While the majority of contemporary quantum processors encode and process information in quantum two-level systems (qubits), processors based on $d$-level \emph{qudits} ($d > 2$) could both (i) store exponentially greater information and (ii) implement certain algorithms using fewer entangling gates than their qubit-based counterparts~\cite{ralph2007efficient,gottesman1998fault,gedik2015computational,PhysRevLett.94.230502}.  Recently, diverse experimental platforms including optical photons, nitrogen-vacancy centers, trapped ions, and superconducting circuits have begun to explore qudit-based information processing~\cite{lanyon_manipulating_2008, naik_random_2017, wang_efficient_2019, luo_quantum_2019, imany_npjQI_2019,Blok2020, dolde_high-fidelity_2014, senko_realization_2015, bianchetti_control_2010}. 

In particular, systems based on three-level qutrits are attracting growing interest. Qutrit-based processors can enable, in theory, error correction with small code size \cite{muralidharan2017overcoming, campbell2014enhanced}, high-fidelity magic state distillation \cite{campbell2012magic}, and robust quantum cryptography \cite{bechmann2000quantum,bruss2002optimal} and communication \cite{vaziri2002experimental} protocols.  Experimentally, single qutrits have both enabled fundamental tests of quantum mechanics~\cite{lapkiewicz_experimental_2011} and been used as auxiliary systems to aid various tasks.  Multi-qutrit algorithms have also been executed recently, both with a measurement-based photonic platforms \cite{luo_quantum_2019} and a superconducting five-qutrit processor~\cite{Blok2020}.  With the number of qutrit-based processors steadily increasing, there is a clear need for quantum characterization, verification, and validation (QCVV) techniques suited to these systems.  

Of the variety of extant QCVV protocols, randomized-benchmarking (RB) ~\cite{Knill2008-rz, Magesan2012-dy} has emerged as a workhorse in the field. In standard RB, a limited set of randomly-chosen gate sequences is run on a quantum processor to characterize the average gate fidelity independent of state-preparation-and-measurement (SPAM) errors.  Building on standard RB, interleaved and cycle benchmarking~\cite{Erhard2019} variants can provide error analyses of specific gates.  Randomized benchmarking is typically used to measure fidelities of single- and two-qubit gates, and but has thus far not been implemented in qutrit systems.  

In this letter, we develop and demonstrate explicit qutrit-capable recipes for both randomized benchmarking and cycle benchmarking, and experimentally demonstrate their viability on a quantum processor comprising superconducting qutrits.  Specifically, we report the use of: (i) the standard RB protocol to measure average gate fidelity over single-qutrit Clifford gates; (ii) interleaved RB to measure the fidelity of individual single-qutrit gates; (iii) cycle benchmarking to characterize a two-qutrit entangling gate; and (iv) simultaneous RB on several qutrits to characterize and mitigate crosstalk.  On our processor, measured single-qutrit randomized benchmarking numbers are as low as $2.38 \times 10^{-3}$, quite close to single-qubit gate errors on the same chip.  Our two-qutrit entangling gate, the controlled-SUM, achieves a process fidelity of $0.82$.

\emph{Processor and gateset}---Our quantum processor, introduced and detailed in~\cite{Blok2020}, is comprised of five superconducting transmon circuits. We operate these as qutrits, encoding information in the lowest three transmon energy levels $\{\ket{0},\ket{1},\ket{2}\}$. Our elementary single-qutrit gateset consists of rotations in both the $\{\ket{0},\ket{1}\}$, and $\{\ket{1},\ket{2}\}$ subspaces.  These subspaces can be selectively addressed with microwave pulses on resonance with their transition frequency. In addition to these microwave gates, we can add phases to any state using virtual gates~\cite{McKay2017} in software.  Our two-qutrit gate is a controlled-SUM gate, the qutrit analog of the CNOT gate~\cite{gottesman1998fault}. Dispersive measurement allows us to resolve, in a single shot, the $\ket{0}$, $\ket{1}$, and $\ket{2}$-state occupancies.

\emph{Qubit-like randomized benchmarking}---To begin, we illustrate the problems associated with simply extending qubit randomized benchmarking to all three two-level subspaces of a qutrit: $\{\ket{0},\ket{1}\}$, $\{\ket{1},\ket{2}\}$, and $\{\ket{0},\ket{2}\}$.  While this ``qubit-like RB'' approach can give some useful characterizations of gates, it suffers from certain drawbacks.   First, this approach will yield one RB number per subspace, which does not allow for easy interpretation, comparison between gate implementations, or comparison with a qubit-based processor.  Further, benchmarking an individual gate via interleaved randomized benchmarking is not possible with this approach. Interleaved RB relies on the use of the Clifford group to map any error into a fully depolarizing channel, which is not true for qubit-like RB.

Figure 1 shows the results of performing this qubit-like RB on our processor.  For each subspace, we generate sequences, of depth 2 to 3000, of uniformly random single-qubit Clifford gates, with an inversion gate appended to the end to make the total sequence equivalent to the identity operation.    The Clifford randomization converts arbitrary gate errors into a depolarizing channel in the relevant subspace, allowing for an extraction of the RB value by measuring the survival probability of the initial state.  This survival probability $P$ follows an exponential decay $P = Ap^m + B$, where $m$ is the circuit depth and $A$, $B$, and $p$ are fit parameters.  From the value of $p$, the average error per Clifford operation can be inferred as $r = (1-p)(d-1)/d$, with $d$ the system dimension.  One subtlety when using this method in a qutrit system is that the decay must be applied to the \emph{renormalized} population, where the population of the non-driven state is excluded.  

    \begin{figure}
        \includegraphics{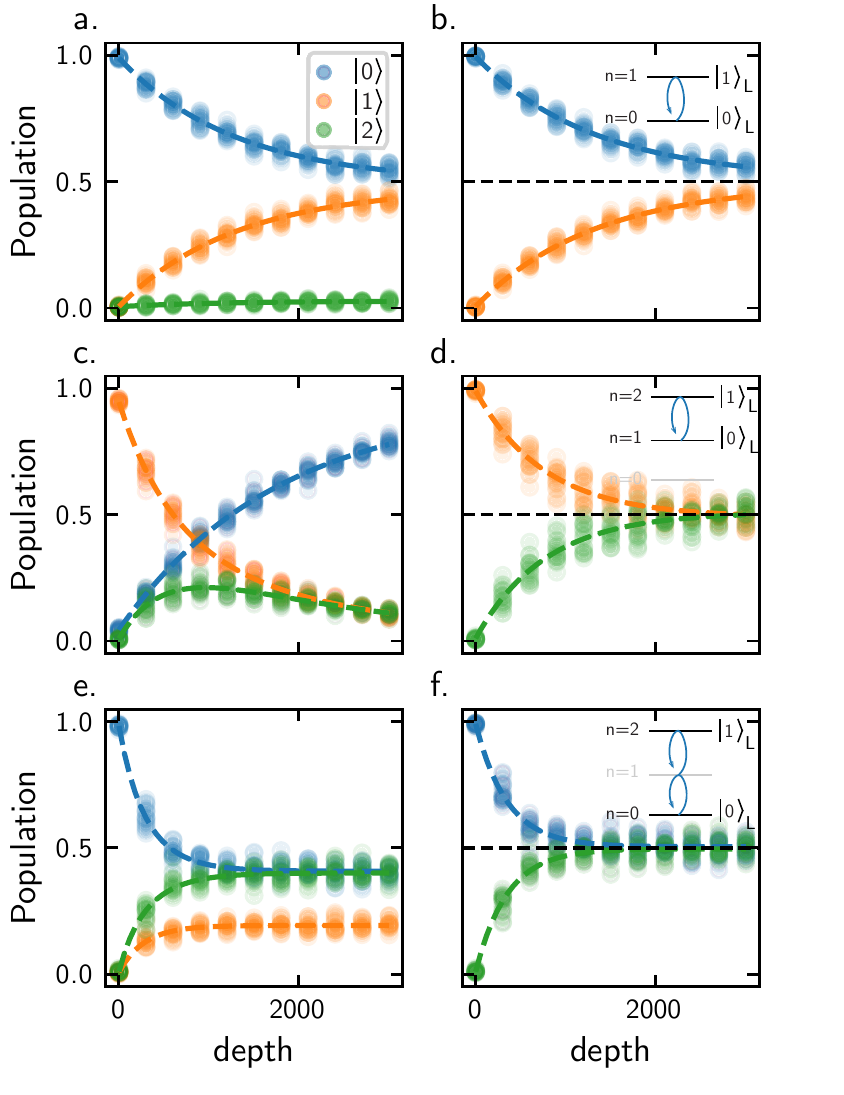}
        \caption{\label{fig:1Q_RB} Population measurement with qubit-like randomized benchmarking. (a-b) RB on the qubit subspace. For a large number of gates, the system tends to a fully depolarized state for the first two levels (b). The second level is nearly unpopulated. (c-d) qubit RB applied on the $\{ \ket{1},\ket{2}\}$ subspace. As the system is depolarizing in this subspace (d), the states decay into the ground state $\ket{0}$. (e-f) qubit RB applied on the on the $\{ \ket{0},\ket{2}\}$. the system tends to a totally mixed state in this subspace with a steady-state population in the state $\ket{1}$.}
    \end{figure}

   Focusing on the randomized-benchmarking results in the $\{\ket{0},\ket{1}\}$ subspace, we find an average infidelity of $r_{\text{qubit}}= (0.39 \pm 0.05) \times 10^{-3}$. Our single-shot readout reveals a small leakage out of the computational space due to off-resonant driving to the $\ket{2}$ state.  This leakage is more pronounced for the other subspaces: for the $\{ \ket{1},\ket{2}\}$ subspace, it is primarily due to $T_1$ decay from $\ket{1}$ to $\ket{0}$, while for the $\{ \ket{0},\ket{2}\}$ subspace, it is primarily due to $T_1$ decay from $\ket{2}$ to $\ket{1}$.  Renormalizing these populations isolates the effects of depolarization within the computational subspace and allows for the extraction of randomized-benchmarking numbers: see Table 1.  This renormalization, however, naturally does not account for leakage errors or dephasing outside the qubit-computational space.  Further, as the curves in Figure 1 show, the qubit RB sequences do not depolarize the full qutrit state, preventing interleaved RB.  To rectify these issues, we turn to genuine qutrit randomized benchmarking. 
    
\emph{Single-qutrit randomized benchmarking}---In order to fully depolarize gate errors in the full qutrit subspace, as required by randomized benchmarking, one needs to run gate sequences in which each gate is sampled from the \emph{qutrit} Clifford group~\cite{Gambetta2012-sf}.  In arbitrary system dimension, the single-qudit Clifford group is defined as the normalizer of the Pauli subgroup, and is generated by the Hadamard gate $H$ and phase gate $S$: \begin{equation}
H = \frac{1}{\sqrt{3}} \begin{pmatrix} 
1 & 1 & 1\\
1 & \omega & \omega^2\\
1 & \omega^2 & \omega
\end{pmatrix}, S = \begin{pmatrix}
1 & 0 & 0\\
0 & 1 & 0\\
0 & 0 & \omega
\end{pmatrix},
\label{eq:matrix}
\end{equation} where $\omega$ is the primitive $d^{th}$ root of unity.  The size of the group grows with system dimension, as $d^3(d^2-1)$ for a single qudit (modulo a global phase); thus the single-qutrit Clifford group has 216 elements.   In principle, any gateset capable of generating the Hadamard and the phase gate can be used to generate the Clifford group.  In our case, we compile Clifford gates into our native gateset using a generalization of the so-called $ZXZXZ$ decomposition for qubits~\cite{McKay2017}---see supplement for details.  On average, our compilation requires 3.325 native gates (not including software-defined phase gates) per qutrit Clifford. 

    \begin{figure}
        \includegraphics{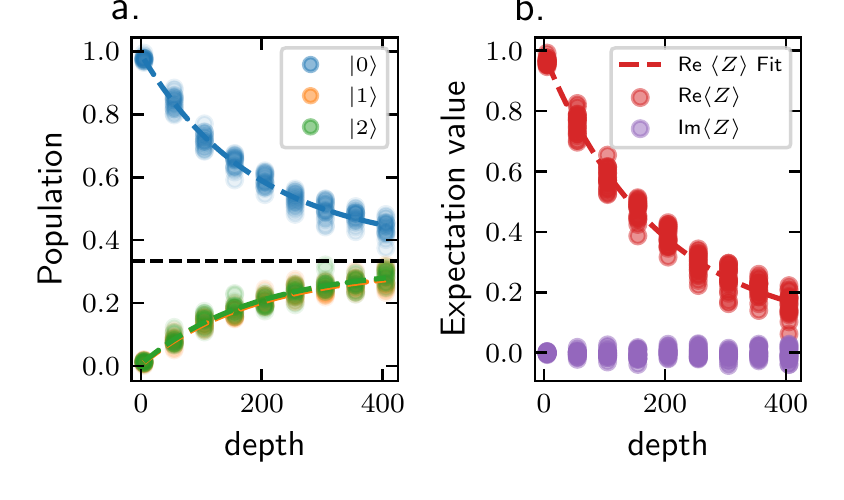}
        \caption{\label{fig:qutrit_RB} Qutrit randomized benchmarking. Panel a. shows the evolution of the population in a randomized benchmarking experiment. With a qutrit Clifford twirling, the noise is effectively mapped into a qutrit depolarizing channel. Thus, for a very long sequence, the populations all tend to $1/3$ and the state of the qubit is on average equal to the maximally mixed state $\rho = \frac{1}{3}\mathbbm{1}_3$. Panel b. illustrates the utility of measuring the expectation value of the Pauli operator $Z$ and taking its real part. The imaginary part of the expectation value remains at zero throughout, which means that the randomized benchmarking sequence is effectively depolarizing the noise.}
    \end{figure}

Using the decomposition, we implement full qutrit randomized benchmarking on our superconducting processor, with results shown in Figure 2.  Figure 2a displays the decay of basis-state populations to their steady-state value; importantly (and in contrast to the results shown in figure 1), the population of all states decays to 1/3, as expected for a fully depolarized state.  This is a signature of Clifford twirling. This depolarization is also illustrated in Figure 2b, which plots the expectation value of the (qutrit) Pauli $Z$ operator.
To extract an RB number, we can fit this decay of $\langle Z \rangle$ to a single exponential
\begin{equation}
    \langle Z \rangle (m) = A p^m + b.
\end{equation}
Note that, in our superconducting processor, measuring $\langle Z \rangle$ can be done with a single circuit due to the fact that we measure all state occupancies at once, allowing for a calculation of 
\begin{equation}
    \langle Z \rangle = P(\ket{0}) + \omega P(\ket{1}) + \omega^2  P(\ket{2}).
\end{equation}
Though the Pauli $Z$ operator is in general non-Hermitian, its phase does not change under depolarization; thus the imaginary component, which by preparation was zero initially, remains zero throughout.  This is plotted for confirmation.  

Using this procedure, we fully characterized the single-qutrit gates of our 5 qutrit chip~\cite{Blok2020}. Table 1 summarizes these results.  Comparing the qutrit error-per-Clifford to the errors measured in qubit randomized benchmarking reveals that the qutrit Clifford gates perform roughly ten times worse than single-qubit Cliffords in the $\{\ket{0},\ket{1}\}$ subspace, and around six times worse than the $\{\ket{1},\ket{2}\}$ subspace.  Part of this difference is explained by the larger number of elementary gates needed to build a single qutrit Clifford than a single qubit Clifford, but this does not explain the full performance gap.  Instead, this performance gap highlights the fact that benchmarking the performance of all our microwave pulses \emph{in the full qutrit Hilbert space} catches errors missed by simply performing single-qubit benchmarking.  Thus, to predict the performance of qutrit algorithms, it is important to run full qutrit randomized benchmarking.

\setlength{\tabcolsep}{0.75em} 
\begin{table*}[htbp]
\begin{tabular}{|c||c|c|c||c|c|c||c|}
\hline
Transmon & \multicolumn{3}{c||}{qubit-like RB: $r \times 10^3$} & \multicolumn{3}{c||}{Leakage: $p_l \times 10^3 $} & qutrit RB $r \times 10^3$ \\
  & $\{ \ket{0},\ket{1}\}$  & $\{ \ket{1},\ket{2}\}$  & $\{ \ket{0},\ket{2}\}$ & $\{ \ket{1},\ket{2}\}$      & $\{ \ket{1},\ket{2}\}$      & $\{ \ket{0},\ket{2}\}$ &           \\ \hhline{|=#=|=|=#=|=|=#=|}
1 & $0.38  \pm 0.01$ & $0.65 \pm 0.03$ & $1.43 \pm 0.05$ & $0.87 \pm 0.16$ & $0.57 \pm 0.02$ & $3.55 \pm 0.21$ & $3.94 \pm 0.15$ \\
2 & $0.36  \pm 0.07$ & $0.38  \pm 0.02$ & $1.01  \pm 0.05$ & $0.99 \pm 0.15$ & $0.22 \pm 0.02$ & $2.98 \pm 0.16$ & $2.38 \pm 0.18$ \\
3 & - & -          & -          & -        &   -      &   -      &  $2.50 \pm 0.21$         \\
4 &  $0.27  \pm 0.02$ & $1.48  \pm 0.11$ & $2.52  \pm 0.22$ & $1.76 \pm 2.49$ & $0.40 \pm 0.02$ & $8.02 \pm 2.40$ & $3.75 \pm 0.59$ \\
5 & $0.55  \pm 0.08$ & $0.63  \pm 0.02$ & $1.18  \pm 0.05$ & $0.87 \pm 0.26$ & $0.28 \pm 0.02$ & $3.87 \pm 0.22$ & $3.34 \pm 0.18$ \\ \hline       
\end{tabular}
    \caption{\label{tab:rb_number} Randomized benchmarking results for the different RB techniques. The parameter $r$ refers to the average error per Clifford $r=\frac{d-1}{d}(1-p)$. The leakage is reported as the exponential decay without any scaling. The reported uncertainties are calculated from the standard deviation of the fit. While doing the qubit-like RB, qutrit 3 became unusable due to a two-level fluctuator poisoning the behavior of the qutrit.}

\end{table*}

We now discuss two applications of single-qutrit RB---characterizing individual gates and measuring crosstalk---before moving on to two-qutrit gates.

\emph{Interleaved randomized benchmarking}---As with qubits, the standard RB protocol measures gate error averaged over the set of Clifford gates.  To characterize individual gates, we turn to the interleaved RB protocol, commonly used in qubit processors.  In this protocol, one interleaves the gate of interest between the Clifford twirls.  For the final inversion gate, it is helpful if the interleaved gate is itself a Clifford gate.  By comparing the rate of depolarization with and without the interleaved gate, a measure of the gate error can be obtained.

Our elementary microwave gates are $\pi$- and $\pi/2$-rotations about the $X$ and $Y$ axes of both the $\{ \ket{0},\ket{1}\}$ and $\{\ket{1},\ket{2}\}$ subspaces. The $\pi$ rotations around the $X$ axis are elements of the single-qutrit Clifford group and can thus be easily characterize by interleaved RB. 

Figure 3 shows the decay curves obtained for our native $\pi$ rotations gates using interleaved RB. The estimation of the gate error is given by~\cite{Magesan2012}:
\begin{equation}
r_{\text{gate}} = \frac{d-1}{d}\left(1-p_{\text{i}}/p\right),     
\end{equation}
where $p_i$ and $p$ are the decay probabilities respectively with and without the interleaving.  This analysis confirms the results of the qubit-like randomized benchmarking: the errors are largest for gates in the $\{\ket{1},\ket{2}\}$ subspace.  It also reinforces the necessity of full qutrit randomized benchmarking: the gate errors measured by qutrit interleaved RB for both $\{\ket{0},\ket{1}\}$ rotations and $\{\ket{1},\ket{2}\}$ rotations are higher than the measured values when only using qubit-like RB. 

In addition to characterizing individual microwave gates, we also characterized the qutrit Hadamard gate.  This composite gate, is one of the Clifford group generators and is used in many qutrit algorithms.  For example, it can be used to implement the quantum Fourier transform in a single qutrit.  For this gate, we report an infidelity of $r_H = (4.6 \pm 0.3) \times 10^{-3}$.  Though our implementation is slow (128 ns), we still achieve a gate fidelity competitive with other implementations of this gate~\cite{Yurtalan2020}.

\begin{figure}[htbp]
    \includegraphics{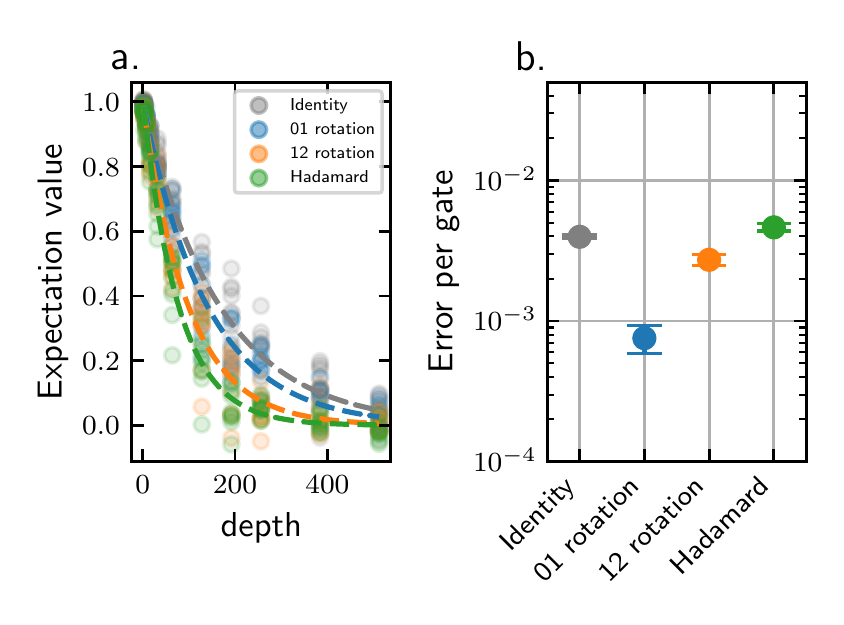} \caption{\label{fig:qutrit_RB)interleaved} Interleaved randomized benchmarking. Panel a shows the depolarization decay for each interleaved gate. Panel b shows the measured error with interleaved benchmarking for several important gates: $\pi$-rotation in the $\{\ket{0},\ket{1}\}$ and $\{\ket{1},\ket{2}\}$ subspace as well as the qutrit Hadamard gate. Note that the gray indicates the interleaved identity - or standard Randomized benchmarking - and is plotted as a reference for the interleaved RB.}
\end{figure}

\begin{figure}[htbp]
        \includegraphics{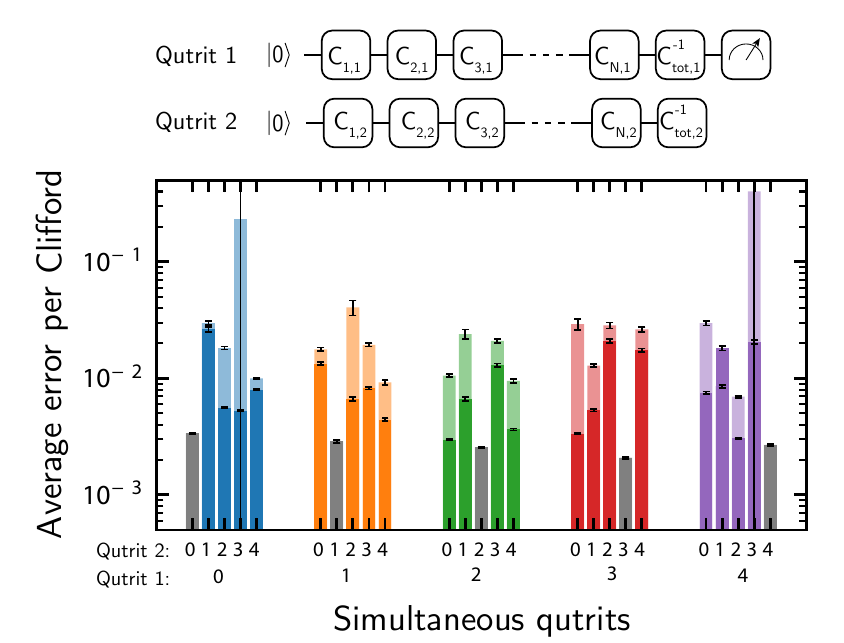}
        \caption{\label{fig:crosstalk_RB} Average error per Clifford while simultaneously running two qutrits. The solid bars indicates the average error per Clifford with the cross-talk nulling applied to the system. The transparent bars indicate error per Clifford without crosstalk nulling. The grey bars indicates the isolated qutrit results. Note that the large uncertainties are for non-cancelled crosstalk .}
    \end{figure}

\emph{Crosstalk}---Another important application of randomized benchmarking is to measure the addressability, or single-qudit gate crosstalk, on a given device \cite{Gambetta2012-sf}.  Usually, crosstalk errors are characterized with a simultaneous randomized-benchmarking experiment where two (or more) qudits undergo an RB sequence at the same time. Transmon qutrits are more sensitive to crosstalk than qubits, as the use of the second level of the transmons increases the frequency crowding of the device.  A significant challenge in the use of a transmon-based device as a qutrit processor is to successfully mitigate these unwanted effects \cite{Blok2020}. Simultaneous qutrit RB allows us to quantify this crosstalk.  Figure 4 shows the the results of the simultaneous RB for two situations: (i) without any crosstalk-cancellation, and (ii) with the nulling protocol developed in~\citet{Blok2020}. Figure~\ref{fig:crosstalk_RB} makes apparent the significant improvement on the crosstalk from such a procedure. Notably, the gain is not homogeneous over the whole chip.
    
\emph{Cycle benchmarking for two-qutrit gates}---We now turn to characterizing two-qutrit gates.  While randomized benchmarking is in principle possible for such a task, it would require sampling from the full two-qutrit Clifford group.  This is impractical both due to the size of the Clifford group (roughly $5\times 10^ 6$ elements) and due to the fact that the vast majority of these gates require multiple elementary entangling gates to implement.  To circumvent these problems and properly characterize two-qutrit entangling gates, we instead generalize the recently-demonstrated technique of cycle benchmarking~\cite{Erhard2019}.

Cycle benchmarking is similar in spirit to interleaved RB in that the gate under test is interleaved between randomly-chosen gates; a key difference is that in cycle benchmarking, the gate of interest is interleaved between Pauli gates rather than Clifford gates.  A key advantage of Pauli twirling over Clifford twirling is that two-qudit Pauli gates are simply tensor products of one-qudit Pauli gates, and thus do not require entangling gates to implement.  Unlike Clifford twirling, however, Pauli twirling does not fully depolarize the gate noise; instead, it maps all errors into stochastic Pauli errors.  Normally, this would result in population survival curves which, unlike RB, do not follow a single exponential decay.  However, in cycle benchmarking, the initial state is chosen to be a Pauli eigenstate, and the final measurement basis is the Pauli basis.  This recovers a single exponential decay.  Averaging over the measured decay parameters for each initial state and measurement basis yields the average error per gate.

The protocol for a qutrit system is quite similar to the qubit protocol for the main twirling part: the gate of interest is interleaved in a random sequence of qutrit Pauli gates. However, several differences appear in the measurement and the state preparation necessary to measure a single exponential decay at a time. The most notable difference comes from the fact that each Pauli operator commutes with its Hermitian conjugate $P^\dagger = P^2$  which is also a Pauli operator. Hence, they share the same eigenbasis and the qutrit case of the cycle benchmarking only uses 4 state preparation and measurement operators $B_Q$ for a single qutrit - compared to 3 for the qubit case - leading to a total of $4^N$ for a $N$-qutrit system for the choice of the initial gate when applying the cycle benchmarking protocol. The four basis rotations are given in the supplementary material and we remark that the qutrit Hadamard takes the computational basis into the eigenbasis of the qutrit Pauli operator $X$ and $X^2$.
Another important point arises from this observation: the expectation values of a Pauli operator and its hermitian conjugate are complex conjugates, \textit{i.e.} $\langle P^\dagger \rangle = \overline{\langle P \rangle}$ implying that the Pauli decay associated with these two operators is the same. A more detailed analysis is given in the supplementary material. This allows us to report the decay associated with only half of all the $N$-Pauli channels. All these specificities come from the structure of the Pauli group and generalize naturally for qudit systems.

With this generalization of the cycle benchmarking protocol to a qutrit system, we have applied it to the identity gate (as a control) and the two-qutrit controlled-SUM gate~\cite{Blok2020}. We measure an average process fidelity of 0.98 for the identity operator and an average process fidelity of 0.82 for the CSUM gate. Using Eq. 4, this corresponds to a gate fidelity of 0.85.

\begin{figure}[htbp]
    \includegraphics{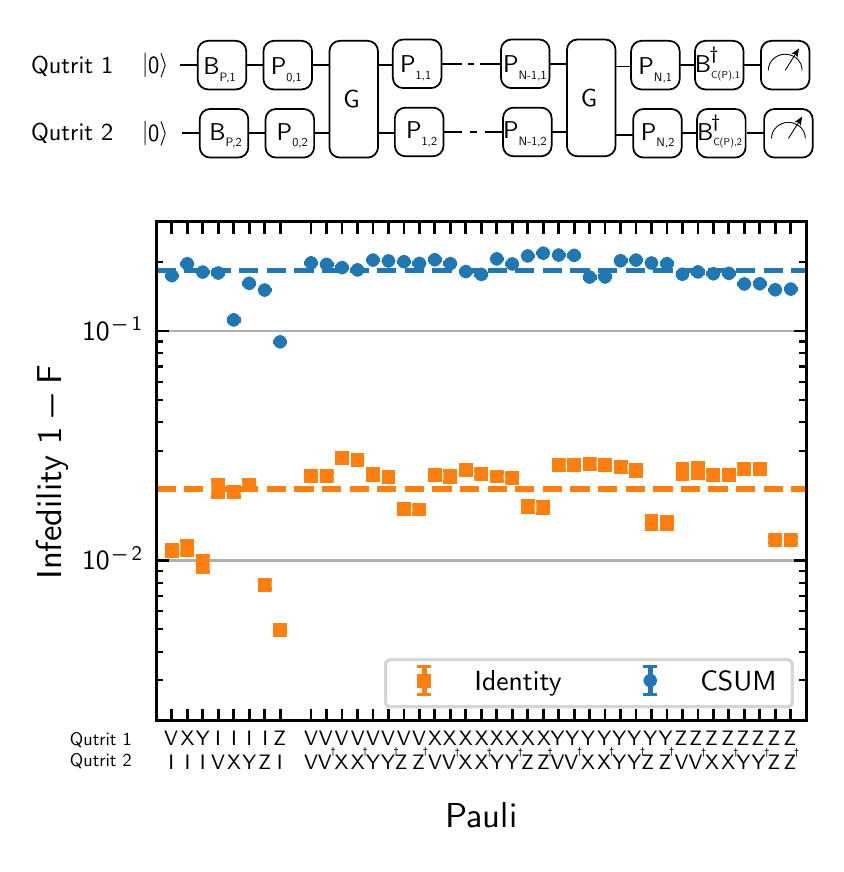}
    \caption{\label{fig:CSUM_CB} Cycle benchmarking of identity two-qutrit cycle (orange square) and of the CSUM cycle (blue circle). We present the exponential decay obtained for each Pauli channel and we have omitted the complex conjugate as its value is identical. The process fidelity is calculated by doing an average over all the Pauli channel and we find a process fidelity of $2.22\times 10^{-2}$ for the identity cycle of $1.83\times 10^{-1}$ for the CSUM.}
\end{figure}

\emph{Conclusion}---We have demonstrated the characterization of a qutrit processor by extending randomized benchmarking and cycle benchmarking to qutrits. With the protocols developed in this article, qutrit processors can be meaningfully compared to their qubit-based counterparts.  As randomized benchmarking has become a workhorse in characterizing qubit processors, we anticipate that qutrit randomized benchmarking will see similarly widespread use.  In addition to developing qutrit randomized benchmarking protocols, we have demonstrated the viability of qutrit processors based on transmon circuits: specifically, single-qutrit gates---both in isolation and simultaneous---achieve fidelities comparable to qubit-based devices.

\vspace{5pt}

\begin{acknowledgments}
    The authors gratefully acknowledge the conversations and insights of D. Gottesman, J. Emerson, J. Wallman, W. Livingston and J.-L. Ville.
    
    This work was supported by the testbed program of the Advanced Scientific Computing Research for Basic Energy Sciences program, Office of Science of the U.S. Department of Energy under Contract No. DE-AC02-05CH11231.
\end{acknowledgments}

\bibliography{main}

\end{document}


\title{Supplementary Material: Qutrit randomized benchmarking}

\author{A. Morvan}
\email{amorvan@lbl.gov}
\affiliation{Quantum Nanoelectronics Laboratory, Dept. of Physics,
             University of California at Berkeley, Berkeley, CA 94720, USA}
\affiliation{Computational Research Division, Lawrence Berkeley National Lab, Berkeley, CA 94720, USA}
\author{V. V. Ramasesh}
\affiliation{Quantum Nanoelectronics Laboratory, Dept. of Physics,
             University of California at Berkeley, Berkeley, CA 94720, USA}
\author{M. S. Blok}
\affiliation{Quantum Nanoelectronics Laboratory, Dept. of Physics,
             University of California at Berkeley, Berkeley, CA 94720, USA}
\affiliation{Department of Physics and Astronomy, University of Rochester, Rochester, NY, 14627 USA}
\author{J.M. Kreikebaum}
\affiliation{Quantum Nanoelectronics Laboratory, Dept. of Physics,
             University of California at Berkeley, Berkeley, CA 94720, USA}
\affiliation{Materials Sciences Division, Lawrence Berkeley National Lab, Berkeley, CA 94720, USA}
\author{K. O'Brien}
\affiliation{Department of Electrical Engineering and Computer Science, Massachusetts Institute of Technology, Cambridge, MA 02139, United States of America}
\author{L. Chen}
\affiliation{Quantum Nanoelectronics Laboratory, Dept. of Physics,
             University of California at Berkeley, Berkeley, CA 94720, USA}
\author{B. K. Mitchell}
\affiliation{Quantum Nanoelectronics Laboratory, Dept. of Physics,
             University of California at Berkeley, Berkeley, CA 94720, USA}
\author{R. K. Naik}
\affiliation{Quantum Nanoelectronics Laboratory, Dept. of Physics,
             University of California at Berkeley, Berkeley, CA 94720, USA}
\author{D. I. Santiago}
\affiliation{Quantum Nanoelectronics Laboratory, Dept. of Physics,
             University of California at Berkeley, Berkeley, CA 94720, USA}
\affiliation{Computational Research Division, Lawrence Berkeley National Lab, Berkeley, CA 94720, USA}
\author{I. Siddiqi}
\affiliation{Quantum Nanoelectronics Laboratory, Dept. of Physics,
             University of California at Berkeley, Berkeley, CA 94720, USA}
\affiliation{Computational Research Division, Lawrence Berkeley National Lab, Berkeley, CA 94720, USA}
\affiliation{Materials Sciences Division, Lawrence Berkeley National Lab, Berkeley, CA 94720, USA}

\date{\today}

\maketitle

\section{Single qutrit gate decomposition}
    Our native gateset for qutrit is given by the two-levels rotations in the subspace spanned by $(\ket{0}, \ket{1})$ and $(\ket{1}, \ket{2})$:
\newcommand{\bigzero}{\mbox{\normalfont\Large\bfseries 0}}
\newcommand{\rvline}{\hspace*{-\arraycolsep}\vline\hspace*{-\arraycolsep}}

    \begin{equation}
        R_{01}(\theta, \phi, \lambda) =
            \begin{pmatrix}
                U(\theta, \phi, \lambda)  &  \rvline & \begin{matrix}
          0 \\
          0
          \end{matrix} \\
        \hline
         \begin{matrix}
          0 &  0 \end{matrix} &  \rvline & 1
        \end{pmatrix}
                \quad \text{and} \quad
                R_{12}(\theta) =
        \begin{pmatrix}
        1 & \rvline &
         \begin{matrix}
              0 &  0 \\ 
         \end{matrix} \\
         \hline
        \begin{matrix}
          0 \\
          0
          \end{matrix} & \rvline & U(\theta, \phi, \lambda)
        \end{pmatrix}
    \end{equation}
    with the subscript ($01$) or ($12$) indicating the subspace of the gate and with the two-level operation given by:
    \begin{equation}
     U(\theta, \phi, \lambda) = 
        \begin{pmatrix}
        \cos \theta/2 & -ie^{i\lambda} \sin \theta/2\\
        -i e^{i\lambda} \sin(\theta/2) & e^{i(\lambda + \phi)}\cos \theta/2\\
        \end{pmatrix}
        \end{equation}
    For each of these two-levels subspace, we use the so-called $ZXZXZ$ decomposition used for qubits \cite{McKay2017} to implement arbitrary rotation in these subspace and can be written as:
    \begin{equation}
        U(\theta, \phi, \lambda) = Z^{(s)}_{\phi - \pi/2} X^{(s)}_{\pi/2} Z^{(s)}_{\pi-\theta} X^{(s)}_{\pi/2} Z^{(s)}_{\lambda-\pi/2}
    \end{equation}
    where the subscript $(s)$ indicate the corresponding subspace. In order to implement arbitrary singe qutrit gate, we use a cos-sin decomposition given in \cite{Dita_2003} where an arbitrary unitary for single qutrit can be implemented by decomposing into subspace rotation as follow:
    \begin{equation}
        U = Z^{(01)}_{\varphi_1}Z^{(12)}_{\varphi_2} Y^{(12)}_{\theta_1} Y^{(01)}_{\theta_2} Z^{(01)}_{\varphi_3}Z^{(12)}_{\varphi_4} Y^{(12)}_{\theta_3} Z^{(01)}_{\varphi_5}Z^{(12)}_{2\varphi_5}
    \end{equation}
    where each $Y$ rotation is actually implemented using the $ZXZXZ$ decomposition described above. We noticed during the writing of the manuscript that one can find a decomposition that use two rotations in the subspace $(01)$ and one rotation in the $(12)$ subspace simply by relabelling the entry of the matrix. 
    
    As an example, here is the decomposition of a qutrit Hadamard:
    \begin{equation}
 H = \frac{1}{\sqrt{3}} \begin{pmatrix} 
1 & 1 & 1\\
1 & \omega & \omega^2\\
1 & \omega^2 & \omega
\end{pmatrix}
=
\underbrace{
\frac{1}{\sqrt{2}} \begin{pmatrix} 
1 & 0 & 0\\
0 & 1 & -1\\
0 & 1 & 1
\end{pmatrix}}_{Y_{\pi/2}^{(21)}}
\underbrace{
\begin{pmatrix} 
\cos \alpha/2 & - \sin \alpha/2 & 0\\
\sin \alpha/2 & \cos \alpha/2 & 0\\
0 & 0 & 1
\end{pmatrix}}_{Y_{\alpha}^{(01)}}
\begin{pmatrix} 
1 & 0 & 0\\
0 & -1 & 0\\
0 & 0 & -i
\end{pmatrix}
\underbrace{
\frac{1}{\sqrt{2}} \begin{pmatrix} 
1 & 0 & 0\\
0 & 1 & -1\\
0 & 1 & 1
\end{pmatrix}}_{Y_{\pi/2}^{(21)}}
\begin{pmatrix} 
1 & 0 & 0\\
0 & 1 & 0\\
0 & 0 & -1
\end{pmatrix}
\end{equation}
With $\alpha = 0.47766$ calculated numerically. $Y_{\alpha}^{(01)}$ can be further decomposed using virtual $Z$ gates:
\begin{equation}
    Y_{\alpha}^{(01)} = X_{\pi/2}^{(01)} Z_{\pi-\alpha}^{(01)} X_{\pi/2}^{(01)} Z_{-\pi}^{(01)}
\end{equation}
The phases (diagonal matrices) can be implemented in software and don't requires physical pulses.

\section{Generation of the single-qutrit Clifford group}
As stated in the main article, the single Clifford group is generated by the Hadamard $H$ gate and a phase gate $S$ \cite{gottesman1998fault}. To generate the group, we start from an empty list and we generate all the element from the generator. If the generated element matrix is already in the list (up to a global phase), we go to the next element. If it is not in the list, we add it to the list. Because the dimension of the Clifford group for a single qudit is known to be $d^3(d^2-1)$, we can stop this search as soon as the number of element in the list is $d^3(d^2-1)$ or 216 for qutrits. This method is a simple brute force search and is already impracticable for the 2-qutrit Clifford group. For each qutrit Clifford gate, we apply the decomposition described in the previous section. When applicable we have simplified successes $\pi/2$ rotations into singles $\pi$ rotation.

We note that $X_{\pi}^{(01)}$ and $X_{\pi}^{(12)}$ are qutrit Clifford gate but $Y_{\pi}^{(01)}$ and $Y_{\pi}^{(12)}$ are not but as the adding a virtual phase don't add error, their error should be similar to the one of the $X_{\pi}^{(s)}$. Moreover the $\pi/2$ rotation along any axis are not in the Clifford group and there is no simple virtual phase that can map them to a Clifford, making interleaving Benchmarking not as straightforward as for the $\pi$-rotation and the Hadamard gate. We leave the question of how to characterize non-Clifford gate for future work

\section{Qutrit Pauli group}
The single qutrit Pauli group is generated by $Z$ and $X$:
\begin{equation}
X =  \begin{pmatrix} 
0 & 1 & 0\\
0 & 0 & 1\\
1 & 0 & 0
\end{pmatrix},
\quad
Z =\begin{pmatrix} 
1 & 0 & 0 \\
0 & \omega & 0\\
0 & 0 & \omega^2
\end{pmatrix}
\end{equation}
we follow the qubit definition and also use $Y = ZX$ and we add the definition of $V = Z^2X$ for convenience. Note that for each of these operators, its complex conjugate and its square are equal: $P^2 = P^\dagger$, \textit{ie} for each element of the group $P^3 = \text{I}$. One corollary of this observation is that each Pauli operator shares the same eigenbasis as its Hermitian conjugate, allowing to measure two expectation values per circuit.

\section{Pauli expectation values measurement}
Following \cite{Erhard2019}, we measure expectation value as following. For a given $N$-qudit Pauli operator $Q$, let $\mathcal{B}_Q$ be the rotation that maps the computational basis to an eigenbasis of $Q$ (\textit{eg} the Hadamard gate for the single-qubit operator $X$). The measurement protocol gives an outcome $z$ after the rotation in the eigenbasis with $\mathcal{B}^{\dagger}_{Q}$. The expectation value of $Q$ can then be expressed as:
\begin{equation}
    \text{Tr}\left[Q\rho\right] = \sum_{z \in \mathbb{Z}^{N}_{3}} \text{Tr}\left[\mathcal{B}_{Q}(\ketbra{z}{z})Q\right] \text{Pr}(z \vert Q)
\end{equation}
which amounts to a weighted sum over the population measurement. For the rotations into the eigenbasis we use $\mathcal{B}_Z = \text{I}$ and:

\begin{equation}
\mathcal{B}_X = H = \frac{1}{\sqrt{3}} \begin{pmatrix} 
1 & 1 & 1\\
1 & \omega & \omega^2\\
1 & \omega^2 & \omega
\end{pmatrix},
\quad
\mathcal{B}_Y = \frac{1}{\sqrt{3}} \begin{pmatrix} 
1 & 1 & \omega \\
1 & \omega & 1\\
\omega & 1 & 1
\end{pmatrix},
\quad
\mathcal{B}_V = \frac{1}{\sqrt{3}} \begin{pmatrix} 
1 & 1 & \omega^2 \\
1 & \omega^2 & 1\\
\omega^2 & 1 & 1
\end{pmatrix}
\end{equation}

\bibliography{main}